\documentclass[12pt]{article}
\pdfoutput=1

\usepackage{amssymb,amsmath,bm,bbm}
\usepackage{cite}
\usepackage{graphicx}
\usepackage{placeins}
\usepackage{longtable,color}

\definecolor{brown}{rgb}{0.6,0.7,0.}
\definecolor{gray}{rgb}{.38,.38,.38}

\def\simge{\mathrel{\rlap{\raise 0.511ex \hbox{$>$}}{\lower 0.511ex \hbox{$\sim$}}}}
\def\simle{\mathrel{\rlap{\raise 0.511ex \hbox{$<$}}{\lower 0.511ex \hbox{$\sim$}}}}
\def\slash#1{\setbox0=\hbox{$#1$}\dimen0=\wd0
      \setbox1=\hbox{/} \dimen1=\wd1 \ifdim\dimen0>\dimen1
      \rlap{\hbox to \dimen0{\hfil/\hfil}} #1                        \else
      \rlap{\hbox to \dimen1{\hfil$#1$\hfil}}
      /   \fi}

\newcommand{\newsection}[1]{\section{#1}\setcounter{equation}{0}}

\newcommand{\lsim}{
\mathrel{\hbox{\rlap{\hbox{\lower4pt\hbox{$\sim$}}}\hbox{$<$}}}}

\newcommand{\gsim}{
\mathrel{\hbox{\rlap{\hbox{\lower4pt\hbox{$\sim$}}}\hbox{$>$}}}}

\newcommand{\tev}{\, {\rm TeV}}
\newcommand{\gev}{\, {\rm GeV}}
\newcommand{\mev}{\, {\rm MeV}}

\allowdisplaybreaks[2]

\newcommand{\be}{\begin{equation}}
\newcommand{\ee}{\end{equation}}
\newcommand{\bea}{\begin{eqnarray}}
\newcommand{\eea}{\end{eqnarray}}
\newcommand{\nn}{\nonumber}
\newcommand{\bi}{\begin{itemize}}
\newcommand{\ei}{\end{itemize}}

\newcommand{\RE}{{\rm Re}}

\def\Br{{\rm Br}}

\begin{document}
\begin{titlepage}
\vspace*{-1.0truecm}

\begin{flushright}
TUM-HEP-764/10\\
\end{flushright}

\vspace{0.4truecm}

\begin{center}
\boldmath

{\Large\textbf{Lepton Flavour Violation in the Presence of a \\[0.3em] Fourth Generation of Quarks and Leptons}}

\unboldmath
\end{center}

\vspace{0.4truecm}

\begin{center}
{\bf Andrzej J.~Buras$^{a,b}$, Bj\"orn Duling$^a$, Thorsten Feldmann$^a$,\\
Tillmann Heidsieck$^a$, Christoph Promberger$^a$
}
\vspace{0.4truecm}

{\footnotesize
 $^a${\sl Physik Department, Technische Universit\"at M\"unchen,
James-Franck-Stra{\ss}e, \\D-85748 Garching, Germany}\vspace{0.2truecm}

 $^b${\sl TUM Institute for Advanced Study, Technische Universit\"at M\"unchen,  Arcisstr.~21,\\ D-80333 M\"unchen, Germany}

}

\end{center}

\vspace{0.4cm}
\begin{abstract}
\noindent
We calculate the rates for the charged lepton flavour violating (LFV) decays $\ell_i\to\ell_j\gamma$, $\tau\to\ell\pi$, $\tau\to\ell\eta^{(\prime)}$, $\mu^-\to e^-e^+e^-$, {the six three-body leptonic decays 
$\tau^-\to \ell_i^-\ell_j^+\ell_k^-$ and the rate for  $\mu-e$ conversion in nuclei} in the Standard Model (SM3) extended by a fourth generation of quarks and leptons (SM4), {assuming that neutrinos are Dirac particles}. 
We also calculate branching ratios for $K_L \to \mu e$, $K_L \to \pi^0\mu e$, $B_{d,s} \to \mu e$, $B_{d,s} \to \tau e$ and $B_{d,s} \to \tau\mu $. 
We find that the pattern of the LFV branching ratios in the SM4 differs significantly from the one encountered in the MSSM, allowing to distinguish these two models with the help of LFV processes in a transparent manner. Also differences with respect to the Littlest Higgs model with T-parity are found. Most importantly
the branching ratios for $\ell_i\to\ell_j\gamma$, $\tau\to\ell\pi$, $\tau\to\ell\eta^{(\prime)}$, $\mu^-\to e^-e^+e^-$, $\tau^-\to e^-e^+e^-$,
 $\tau^-\to \mu^-\mu^+\mu^-$, $\tau^-\to e^-\mu^+\mu^-$ and 
$\tau^-\to \mu^-e^+e^-$ can all still be as large as the present experimental 
upper bounds but not necessarily simultaneously.
Also the rate for  $\mu-e$ conversion in nuclei can reach the corresponding 
upper bound.
\end{abstract}
\end{titlepage}

\section{Introduction}
It is well known that in the Standard Model with three generations of leptons (SM3),
 the FCNC processes in the charged lepton sector, like $\ell_i\to\ell_j\gamma$ and $\mu^-\to e^-e^+e^-$, 
are very strongly suppressed due to tiny neutrino masses. 
In particular, the branching ratio for $\mu\to e\gamma$ in the SM3 {amounts} to at most $10^{-54}$, 
to be compared with the present experimental upper bound $1.2\cdot 10^{-11}$ \cite{Brooks:1999pu}, 
and with the one that should be available in the coming years, $\sim10^{-13}$ \cite{Yamada:2005tg,psi:2010wp}. 
Therefore any observation of lepton flavour violation (LFV) in the foreseeable future would be an unambiguous signal of 
New Physics (NP) beyond the SM3.

Indeed, in extensions of the SM3 like the MSSM, the Littlest Higgs model with T-parity (LHT) and Randall-Sundrum (RS) models, 
the presence of heavy leptons allows for much higher branching ratios for LFV processes, provided the mixing of these new leptons with the ordinary leptons is sufficiently large.
On the other hand, none of these processes has been observed in experiment so far, and therefore we have to conclude that LFV rates are strongly suppressed in nature. As a consequence, these observations often put severe constraints on the parameter space of NP models in which heavy leptons are present. Moreover, these decays, being unaffected by hadronic uncertainties, allow for a clear distinction between different NP scenarios, in particular when several branching ratios are considered simultaneously
 and patterns of LFV in said extensions are compared to each other. For instance, a clear distinction between {the MSSM} and the 
LHT model \cite{Blanke:2007db} should be possible in this manner, once the relevant experimental data will be available.

Models like the MSSM, LHT and RS scenarios contain many free parameters, and consequently do not always allow for clear-cut conclusions. 
On the other hand, one of the simplest extensions of the SM3, which contains heavy leptons and which has much less free parameters than the 
MSSM, LHT and RS scenarios, is the addition of a sequential fourth generation (4G) of quarks and leptons (SM4).

In two recent papers \cite{Buras:2010pi,Buras:2010nd}, we have analysed FCNC processes in the quark sector within the SM4, 
finding often sizeable departures from the SM3 predictions for a multitude of rare decays of mesons and for 
$K^0-\bar K^0$, $B_{d,s}^0-\bar B_{d,s}^0$, $D^0-\bar D^0$ particle-antiparticle mixings, including in particular CP-violating observables.
For recent work on SM4, see, for instance, \cite{Kribs:2007nz,Soni:2008bc,Holdom:2009rf, Eilam:2009hz,BarShalom:2010bh,Soni:2010xh,Hou:2010mm,Choudhury:2010ya,Chanowitz:2010bm} and references in \cite{Buras:2010pi,Buras:2010nd}.

The goal of the present paper is a new analysis of the $\ell_i\to\ell_j\gamma$, $\ell_i\to3\ell_k$ decays and of other LFV processes 
within the SM4, with the aim of finding the pattern of LFV in this model and constraining the masses of the new leptons and their mixing 
with the ordinary SM3 leptons.
There have been several analyses of LFV within the SM4 in the past \cite{Lacker:2010zz,Huo:2003pw,Hou:2008yb}, but to our knowledge,
 no detailed analysis of correlations between several LFV branching ratios has been presented to date. 
As we will demonstrate below, these correlations could provide us with a very valuable tool for distinguishing the SM4 from other NP scenarios.

Our paper is organised as follows. In Section \ref{sec:mixing-matrix} we provide the $4\times 4$ leptonic mixing matrix $U_{\rm SM4}$ and 
discuss its various properties, taking into account the present experimental information about the $3\times 3$ PMNS matrix. 
{ The subsequent section can be considered as a compendium of formulae for the most interesting branching ratios for LFV processes within the SM4. 
In Section~\ref{sec:mu-e-gamma}, formulae for the dipole transitions $\mu\to e\gamma$, $\tau\to\mu\gamma$ and $\tau\to e\gamma$ are presented. The corresponding 
formulae for the three types of four-lepton transitions $\ell_i^- \to \ell_j^- \ell_j^+ \ell_j^-$, $\tau \to \ell_i^- \ell_j^+ \ell_i^-$ and $\tau^- \to \ell_i^- \ell_j^+ \ell_j^-$ are presented in Section~\ref{sec:four-lepton}. 
In Section~\ref{sec:tau-mu-P} we turn our attention to the semi-leptonic $\tau$-decays, $\tau\to\mu P$ with $P=\pi,\eta,\eta^\prime$,
 and in Section~\ref{sec:mu-e-conversion} we calculate the $\mu-e$ conversion rate in nuclei. 
The Sections~\ref{sec:K-mu-e} and \ref{sec:B-mu-e} deal with 
the decays $K \to (\pi^0) \mu e$ and $B_{d,s} \to 2 \ell$, respectively, while in Section~\ref{sec:g-2} the issue of $(g-2)_\mu$ is discussed.
In Section~\ref{sec:numerics} we present a detailed numerical analysis of all processes listed above, paying particular attention to several correlations. 
As demonstrated in Section~\ref{sec:distinction}, these correlations allow for a clear distinction of the SM4 from the MSSM and the LHT model. 
Finally, in Section~\ref{sec:conclusions} we conclude our paper with a list of messages from our analysis and with a brief outlook. 
Few technical details are relegated to the appendix. }


\boldmath
\section{The $4\times4$ Leptonic Mixing Matrix $U_\text{SM4}$\label{sec:mixing-matrix}}
\unboldmath

The most general leptonic mixing matrix in the 4G case contains 
$$
 \mbox{6 mixing angles}\,, \qquad 
 \mbox{3 Dirac phases}\,, \qquad 
 \mbox{3 Majorana phases}.
$$
A standard parametrisation is obtained by treating the mixing angles and Dirac phases
in analogy to the quark sector (see e.g.\ \cite{Bobrowski:2009ng,Buras:2010pi}; for early work see 
 \cite{Fritzsch:1986gv,Harari:1986xf,Anselm:1985rw}), 
with the Majorana phases contained in an additional diagonal matrix,
\begin{equation}
U_{\rm SM4} =
U_{34} \, I_4(\delta_{24}) \, U_{24} \, I_4(-\delta_{24}) \,
I_4(\delta_{14})\, U_{14} \, I_4(-\delta_{14}) \,
U_{23} \, I_3(\delta_{13}) \, U_{13} \, I_3(-\delta_{13}) \, U_{12}
\, I_{\rm Maj.}
\,,
\end{equation}
where $U_{ij}$ are rotations in the $i$-$j$ plane, parameterised by corresponding mixing angles $\theta_{ij}$, and
\begin{align}
  [I_{i}(\alpha)]_{jk} & = \delta_{jk} \, e^{i \alpha \, \delta_{ij} }\,, 
\qquad 
  I_{\rm Maj.} =  {\rm diag}[e^{i\alpha_1},e^{i\alpha_2},e^{i\alpha_3},1] \,,
\end{align}
contain the Dirac and Majorana phases, respectively.
The observables we are going to consider in the following are insensitive to the Majorana phases $\alpha_i$,
which will therefore be dropped in the following.

Concerning the mixing angles and Dirac phases, it is well known that the SM3 lepton sector behaves
very differently as compared to the SM3 quark sector.
In particular, the PMNS matrix for SM3 leptons is known to follow an approximate
``tri-bi-maximal'' mixing pattern \cite{Harrison:2002er}, where
\begin{align}
 U_{\rm SM3} &\simeq U^{\rm max}_{\rm SM3} =
\begin{array}{cc}
 &  \begin{array}{ccc}
       \nu_1 \quad & \nu_2 & \quad \nu_3
      \end{array}
\\[0.3em]
 \begin{array}{c}
 e \\ \mu \\ \tau
\end{array} &
\left(
\begin{array}{ccc}
 \sqrt{\frac{2}{3}} & \frac{1}{\sqrt{3}} & 0 \\
 -\frac{1}{\sqrt{6}} & \frac{1}{\sqrt{3}} & \frac{1}{\sqrt{2}} \\
 \frac{1}{\sqrt{6}} & -\frac{1}{\sqrt{3}} & \frac{1}{\sqrt{2}}
 \end{array}
 \right)
\end{array}\,,
\end{align}
which corresponds to the situation where the 3G mixing angles satisfy 
$s_{23} =1/\sqrt{2}$, $s_{12}=1/\sqrt{3}$, $s_{13}=0$.

A priori, it is not clear whether such a pattern could or should be
extended to an SM4 lepton-mixing matrix, leading to potentially 
large mixing angles between the new lepton generation and the SM3 ones.
However, as we will see in more detail below, the current experimental
situation already excludes large new mixing angles $\theta_{i4}$ with
the 4G leptons, and therefore we should rather consider
\begin{align} U_{\rm SM4} \approx   U^{\rm max}_{\rm SM4} & =  
\left(
\begin{array}{cccc}
 \sqrt{\frac{2}{3}} & \frac{1}{\sqrt{3}} & 0 & 0\\
 -\frac{1}{\sqrt{6}} & \frac{1}{\sqrt{3}} & \frac{1}{\sqrt{2}} & 0 \\
 \frac{1}{\sqrt{6}} & -\frac{1}{\sqrt{3}} & \frac{1}{\sqrt{2}} & 0 \\
 0 & 0 & 0 & 1 
 \end{array}
 \right) 
\label{newpar}
\end{align}
as a starting point.\footnote{This ansatz reflects the special role of the fourth-generation neutrino, which
requires some particular theoretical framework to be realized (see e.g.\ \cite{Burdman:2009ih}).}
The \emph{deviations} from this mixing pattern can then be
conveniently described in terms of an almost diagonal mixing matrix, 
\begin{align}
 U_{\rm SM4} &= V^{\rm residual}_{\rm SM4} \cdot U^{\rm max}_{\rm SM4} \cdot I^{\rm Maj.} \,,
\end{align}
where $V^{\rm residual}$ is parameterised in terms of small mixing angles $\Delta_{ij}$ 
and 3 Dirac phases, and can be treated in an analogous way as the 4G quark mixing matrix. 

In particular, as in the quark sector, we may require the mixing angles $\Delta_{ij}$ 
to fulfill consistency relations \cite{Buras:2010pi,Feldmann:2006jk},
\begin{align}
 \Delta_{ik} \Delta_{jk} \lesssim \Delta_{ij} \qquad \mbox{(no summation over $k$)} \,.
\end{align}
For $k=4$, the product $\Delta_{i4} \Delta_{j4} \sim |U_{i4} \,
U_{4j}^*|$
on the left-hand side of this relation determines, for instance,
the size of radiative $\ell_i \to \ell_j$ decays (see next section),
which in turn set a lower bound (order-of-magnitude-wise) on the
deviations of
the PMNS matrix from tri-bi-maximal mixing, with
\begin{align}
& U_{e2}  \approx \frac{1 + \Delta_{12}-\Delta_{13}}{\sqrt3} \,, \quad
 U_{e3}  \approx \frac{\Delta_{12}+\Delta_{13}}{\sqrt2}\,, \quad
 U_{\mu3}  \approx \frac{1+\Delta_{23}}{\sqrt2} \,.
\end{align}
While the deviations from tri-bi-maximal mixing $\Delta_{ij}$  -- with the present experimental bounds -- can still be of order $10-20\%$, the radiative LFV decays (see below) constrain the products
\begin{eqnarray}
\Delta_{14}\Delta_{24} &\simle& 3.5\times 10^{-4}\,,\nonumber\\
\Delta_{14}\Delta_{34} &\simle& 1.8\times 10^{-2}\,,\nonumber\\
\Delta_{24}\Delta_{34} &\simle& 1.8\times 10^{-2}\,.
\end{eqnarray}

\section{Compendium for LFV in the SM4\label{sec:compendium}}
\boldmath
\subsection{Dipole Transitions}\label{sec:mu-e-gamma}
\begin{figure}
\center{
\includegraphics{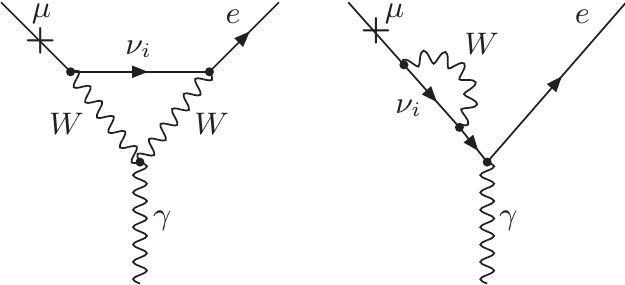}
}
\caption{Diagrams contributing to the $\mu\to e\gamma$ decay.\label{fig:megSM}}
\end{figure}

\unboldmath
The diagrams for $\mu\to e\gamma$ in the SM3 and SM4 are shown in Fig.~\ref{fig:megSM}, and analogous diagrams exist for $\tau\to\mu\gamma$ and $\tau\to e\gamma$. 
As demonstrated in \cite{Blanke:2007db}, the relevant branching ratios can be found by inspecting the analogous calculation for $B\to X_s\gamma$. 
Neglecting tiny contributions from ordinary neutrinos and using formulae (3.8), (3.20) and (3.21) in \cite{Blanke:2007db}, we easily find
\begin{align}
\label{eq:meg}
&\Br(\mu\to e\gamma)=\frac{3\alpha}{2\pi}\left|\chi_4^{(\mu e)} \right|^2H(y_4)^2\,,\\
\label{eq:teg}
&\Br(\tau\to e\gamma)=\frac{3\alpha}{2\pi}\Br(\tau^-\to\nu_\tau e^-\bar\nu_e)\left|\chi_4^{(\tau e)}\right|^2 H(y_4)^2\,,\\
\label{eq:tmg}
&\Br(\tau\to \mu\gamma)=\frac{3\alpha}{2\pi}\Br(\tau^-\to\nu_\tau \mu^-\bar\nu_\mu)\left|\chi_4^{(\tau \mu)}\right|^2 H(y_4)^2\,,
\end{align}
where we have defined $y_j=m_{\nu_j}^2/M_W^2$ and
\be \label{eq:chi}
\chi_j^{(\mu e)} = U_{ej} U_{\mu j}^\ast\,,\quad \chi_j^{(\tau e)} = U_{ej} U_{\tau j}^\ast\,,\quad \chi_j^{(\tau \mu)} = U_{\mu j} U_{\tau j}^\ast\,.
\ee
The involved branching ratios of leptonic $\tau$ decays are \cite{Amsler:2008zzb},
\begin{align}
\label{eq:teg-exp}
&\Br(\tau^-\to\nu_\tau e^-\bar\nu_e)=(17.84\pm0.05)\%\,,\\
\label{eq:tmg-exp}
&\Br(\tau^-\to\nu_\tau \mu^-\bar\nu_\mu)=(17.36\pm0.05)\%\,,
\end{align}
and the loop function
\begin{equation}
H(x)=D_0^\prime(x)-\frac{2}{3}E_0^\prime(x)\,,
\label{eq:H(x)}
\end{equation}
is given in terms of functions $D_0^\prime$ and $E_0^\prime$, known from the analysis of the $B\to X_s\gamma$ decay, see Appendix~\ref{app:functions}.

In writing (\ref{eq:meg})--(\ref{eq:tmg}), we have neglected electroweak (EW) corrections and non-unitarity corrections in the 
PMNS matrix that affect the leptonic decays and have been recently discussed by Lacker and Menzel \cite{Lacker:2010zz}.
Similarly, we have neglected corrections $\sim\mathcal O(m_\mu^2/m_\tau^2)$, $\sim\mathcal O(m_e^2/m_\tau^2)$ and $\sim\mathcal O(m_e^2/m_\mu^2)$. 
These corrections amount to at most a few percent and would only be necessary in the presence of very accurate experimental branching ratios. 

The important virtue of formulae (\ref{eq:meg})--(\ref{eq:tmg}) is that, taken together, they allow for a direct determination of the 
ratios of the elements $|U_{e4}|,\,|U_{\mu4}|$ and $|U_{\tau4}|$, independently of the mass $m_{\nu_4}$. In particular, we have
\begin{align}
&\frac{\Br(\tau\to\mu\gamma)}{\Br(\mu\to e\gamma)}=\left|\frac{U_{\tau4}}{U_{e4}}\right|^2\Br(\tau^-\to\nu_\tau\mu^-\bar\nu_\mu)\,,\\
&\frac{\Br(\tau\to\mu\gamma)}{\Br(\tau\to e\gamma)}=\left|\frac{U_{\mu4}}{U_{e4}}\right|^2\frac{\Br(\tau^-\to\nu_\tau\mu^-\bar\nu_\mu)}{\Br(\tau^-\to\nu_\tau e^-\bar\nu_e)}\approx\left|\frac{U_{\mu4}}{U_{e4}}\right|^2\,,\\
&\frac{\Br(\tau\to e\gamma)}{\Br(\mu\to e\gamma)}=\left|\frac{U_{\tau4}}{U_{\mu4}}\right|^2\Br(\tau^-\to\nu_\tau e^-\bar\nu_e)\,.
\end{align}
For precision measurements, also the higher order corrections have to be included.

Finally, let us note that in contrast to \cite{Lacker:2010zz}, where the formulae of Altarelli et al.~\cite{Altarelli:1977zq} were used, 
we are able to give the results (\ref{eq:meg})--(\ref{eq:tmg}) in an analytic form without any phase space integrals to be evaluated.

\boldmath
\subsection{Four-Lepton Transitions}\label{sec:four-lepton}
\unboldmath
\boldmath
\subsubsection{The Decays $\mu^-\to e^-e^+e^-$, $\tau^-\to\mu^-\mu^+\mu^-$ and $\tau^-\to e^-e^+e^-$\label{sec:mu-3e}}
\unboldmath
Next, we will consider decays with three leptons in the final state. 
Dipole operators, photon penguins, Z-penguins and box diagrams 
contribute here.
In general, this will generate a rather non-trivial Dalitz distribution 
for the final state \cite{Dassinger:2007ru} which
should be taken into account in the experimental analysis. In the SM4 
analysis, we find that the dipole operators are generally sub-leading,
and therefore the Dalitz distributions are rather flat functions of the 
invariant masses of the final-state lepton pairs. It is therefore
sufficient to consider the formulae for the (partially) integrated 
branching ratios, which can be directly obtained from the
corresponding expressions in \cite{Blanke:2007db} by appropriately 
replacing the loop functions
by those of the SM4. We first find, using (5.8) of \cite{Blanke:2007db},
\begin{eqnarray}
\label{eq:meee}
\Br(\mu^-\to e^-e^+e^-) &=& \frac{\Gamma(\mu^-\to e^-e^+e^-)}{\Gamma(\mu^-\to e^-\bar\nu_e\nu_\mu)} \nn\\
&=& \frac{\alpha^2}{\pi^2}\left|\chi_4^{(\mu e)}\right|^2 \bigg[ 3 \bar Z ^2 + 3\,\bar Z H(y_4) + H(y_4)^2\left(\log\frac{m_\mu}{m_e}-\frac{11}{8}\right)\nn\\
&&\qquad +\frac{1}{2\sin^4\theta_W} \bar Y_{e} ^2-\frac{2}{\sin^2\theta_W}\bar Z\bar Y_{e}-\frac{1}{\sin^2\theta_W} H(y_4)\bar Y^{}_{e}\bigg]\,.
\end{eqnarray}
The one-loop functions entering this formula are easily found:
\begin{align}
\label{eq:Ybar}
&\bar Y_e^{}=Y_0(y_4)-|U_{e4}|^2S_0(y_4)\,,\\
\label{eq:Zbar}
&\bar Z^{}=C_0(y_4)+\frac{1}{4}G_0(y_4)\,,
\end{align}
where
\begin{equation}
G_0(x)=D_0(x)-\frac{2}{3}E_0(x)\,.
\end{equation}
The expressions for $Y_0,\,S_0,\,C_0,\,D_0$ and $E_0$ can be found in Appendix~\ref{app:functions}, and $H(x)$ is given in (\ref{eq:H(x)}).
Note that $D_0(x)$ and $E_0(x)$ diverge logarithmically for $x\to0$, but $G_0(x)$ vanishes for $x \to 0$. Since also $C_0(0)=0$, the contributions of light neutrinos can be neglected.

For $\tau^-\to\mu^-\mu^+\mu^-$, we make the following replacements in (\ref{eq:meee}),
\begin{equation}
\chi_4^{(\mu e)}\to \chi_4^{(\tau \mu)}\,,\qquad \bar Y_e\to \bar Y_\mu\,,
\end{equation}
appropriately adjust the charged lepton masses, and multiply (\ref{eq:meee}) by $\Br(\tau^-\to\nu_\tau\mu^-\bar\nu_\mu)$. 
The function $\bar Y_\mu$ can be obtained from $\bar Y_e$ in (\ref{eq:Ybar}) by replacing $|U_{e4}|^2\to |U_{\mu4}|^2$.
In the case of $\tau^-\to e^-e^+e^-$, the corresponding replacements with respect to (\ref{eq:meee}) are
\begin{equation}
\chi_4^{(\mu e)}\to \chi_4^{(\tau e)}\,,
\end{equation}
and again the appropriate charged lepton masses have to be used. 
Now $\Br(\tau^-\to \nu_\tau e^-\bar\nu_e)$ enters the formula for $\Br(\tau^-\to e^-e^+e^-)$ as an overall factor.

\boldmath
 \subsubsection{The Decays $\tau^- \to e^-\mu^+ e^-$ and $\tau^- \to \mu^- e^+ \mu^-$ \label{sec:tau-emue}}
\unboldmath
Again following \cite{Blanke:2007db} and adjusting the formulae given there to the SM4, we find
\be\label{eq:Brteme}
\Br(\tau^-\to e^-\mu^+e^-)=\frac{m_\tau^5\tau_\tau}{192\pi^3}\left(\frac{G_F^2 M_{W}^2}{4\pi^2}\right)^2
\left|\sum_{i,j} \chi_i^{(\tau e)}\chi_j^{(\mu e)} P(y_i,y_j)\right|^2,
\ee
where summation goes over all four neutrinos. {Using unitarity of the 4G leptonic mixing matrix and the fact that only the $i,j=4$ contribution is relevant, (\ref{eq:Brteme}) simplifies to}
\begin{equation}
\Br(\tau^-\to e^-\mu^+e^-)=\frac{m_\tau^5\tau_\tau}{192\pi^3}\left(\frac{G_F^2 M_{W}^2}{4\pi^2}\right)^2
\left|\chi_4^{(\tau e)}\chi_4^{(\mu e)}\right|^2 R_2(y_4)^2\,,
\end{equation}
{where we have introduced the function}
\begin{equation}
R_2(x)\equiv R(x,0,0,x) = 1-2 P(x,0)+P(x,x)\,.
\label{eq:R2}
\end{equation}

The branching ratio for $\tau^-\to\mu^-e^+\mu^-$ is obtained from \eqref{eq:Brteme} by interchanging $\mu\leftrightarrow e$ and using $\chi_j^{(e\mu)}=\chi_j^{(\mu e)\ast}$.

\boldmath
\subsubsection{The Decays $\tau^-\to\mu^-e^+e^-$ and $\tau^-\to e^-\mu^+\mu^-$}
\unboldmath\label{sec:tmee}
Adapting the formulae of \cite{Blanke:2007db} to the SM4, we obtain
\be\label{eq:Brtmee_int}
\Br(\tau^-\to\mu^-e^+e^-)=\Br(\tau^-\to \mu^-\bar\nu_\mu\nu_\tau)\int_{4m_e^2/m_\tau^2}^1 R^{\tau\mu}(\hat s)\,d\hat s\,,
\ee
with the differential decay rate $R^{\tau\mu}(\hat s)$
\bea\label{eq:Rs}
R^{\tau\mu}(\hat s)&=&\frac{\alpha^2}{4\pi^2}(1-\hat s)^2
\bigg[(1+2\hat s)\left(|\tilde{C}_9^{\tau\mu}|^2+|\tilde C_{10}^{\tau\mu}|^2\right)\nn\\
&& \hspace{2.5cm}
+4\left(1+\frac{2}{\hat s}\right)|C_7^{\tau\mu}|^2+12\, \text{Re}(C_7^{\tau\mu} \tilde{C}_9^{\tau\mu\ast})
\bigg]\,.
\eea
The relevant Wilson coefficients read
\begin{gather}
\label{eq:C7}
 C_7^{\tau\mu} = -\frac{1}{2} H(y_4) \chi_4^{(\tau\mu)}\,,\\
\label{eq:C9}
\tilde C_9^{\tau\mu}=\frac{\bar Y_{e} \chi_4^{(\tau\mu)}}{\sin^2\theta_W}-4\bar Z \chi_4^{(\tau\mu)}-\Delta_{\tau\mu}\,,\qquad
\tilde C_{10}^{\tau\mu}=-\frac{\bar Y_{e} \chi_4^{(\tau\mu)}}{\sin^2\theta_W}+\Delta_{\tau\mu}\,,
\end{gather}
where we have introduced the contribution due to additional box diagrams
\begin{eqnarray}
 \Delta_{\tau\mu} &=& \frac{1}{4 \sin^2\!\theta_W} \sum_{i,j} \chi_i^{(\tau e)}{\chi_j^{(\mu e)}}^\ast P(y_i,y_j)\nonumber\\
&=&\frac{1}{4 \sin^2\!\theta_W}\chi_4^{(\tau e)}{\chi_4^{(\mu e)}}^\ast R_2(y_4)\,.
\end{eqnarray}
{In the above we neglected RGE running of $\alpha$ as well as operator mixing. The integral in (\ref{eq:Brtmee_int}) can then be performed analytically, and we arrive at
\begin{eqnarray}\label{eq:Brtmee}
 \frac{\Br(\tau^-\to\mu^-e^+e^-)}{\Br(\tau^-\to \mu^-\bar\nu_\mu\nu_\tau)}= \frac{\alpha ^2}{24 \pi ^2} &\Biggl[& 3 \left(|\tilde C_{10}^{\tau \mu}|^2+|\tilde C_9^{\tau\mu}|^2\right) \left(1- z\right)^3\! \left(1+ z\right)\nonumber\\
&& -8\, |C_7^{\tau\mu}|^2 \left(1- z\right)\left(8-z-z^2\right) \\
&& +24\, \RE(C_7^{\tau\mu} \tilde C_9^{\tau\mu\ast}) \left(1- z\right)^3 - 48\, |C_7^{\tau\mu}|^2 \log \left( z\right) \Biggr]\,,\nonumber
\end{eqnarray}
where we have introduced 
\be z \equiv \frac{4 m_e^2}{m_\tau^2}\,.\ee
}

\noindent Again, the formula for $\Br(\tau^- \to e^- \mu^+ \mu^-)$ is obtained from its analogous counterpart by interchanging $\mu \leftrightarrow e$ in (\ref{eq:Brtmee}).

\boldmath
\subsection{Semi-Leptonic $\tau$-Decays\label{sec:tau-mu-P}} 
\unboldmath
The upper limits from Belle and Babar for the branching ratios of the decays $\tau\to\mu\pi$, $\tau\to\mu\eta$ and $\tau\to\mu\eta^\prime$ are given in Table~\ref{tab:upper-bounds}. 
Analytic expressions for these can be obtained directly from the corresponding formulae (4.12) and (4.17) 
in \cite{Blanke:2007db} by replacing the loop functions of the LHT model by the loop functions of the SM4. Neglecting suppressed pion and muon mass contributions of order
$\mathcal{O}(m_\pi^2/m_\tau^2)$ and $\mathcal{O}(m_\mu^2/m_\tau^2)$, we find
\begin{align}
\label{eq:tmpi}
&\Br(\tau\to\mu\pi)=\frac{G_F^2 \alpha^2 F_\pi^2 m_\tau^3 \tau_\tau}{128\pi^3 \sin^4
  \theta_W }\left|\chi_4^{(\tau\mu)}\right|^2\, (\bar X + \bar Y)^2\,,
\end{align}
with $\tau_\tau$ and $m_\tau$ being the lifetime and mass of the decaying
$\tau$ lepton.
The branching ratio for the $\tau\to e\pi$ decay can  be obtained very easily from (\ref{eq:tmpi}) by simply replacing $U_{\mu4}$ with $U_{e4}$.
The generalisation of (\ref{eq:tmpi}) to the decays $\tau\to\mu\eta$ and
$\tau\to\mu \eta^\prime$ is slightly complicated
by mixing in the $\eta - \eta^\prime$ mesonsystem and has been discussed in detail in \cite{Blanke:2007db}. Proceeding as there, one obtains
\begin{align}
\label{eq:tmeta}
&\Br(\tau\to\mu\eta)=\frac{G_F^2 \alpha^2 F_\pi^2 m_\tau^3 \tau_\tau}{128\pi^3 \sin^4
  \theta_W }\left|\chi_4^{(\tau \mu)}\right|^2\, \left(\frac{\cos \theta_8}{\sqrt{3}}\frac{F_8}{F_\pi}(\bar X^{} + \bar Y^{})
- \sqrt{\frac{2}{3}}\sin \theta_0\frac{F_0}{F_\pi}(\bar X^{}  -2\, \bar Y^{})\right)^2\,,\\
\label{eq:tmetaprime}
&\Br(\tau\to\mu\eta^\prime)=\frac{G_F^2 \alpha^2 F_\pi^2 m_\tau^3 \tau_\tau}{128\pi^3 \sin^4
  \theta_W }\left|\chi_4^{(\tau \mu)}\right|^2\, \left( \frac{\sin \theta_8}{\sqrt{3}}\frac{F_8}{F_\pi}(\bar X^{} + \bar Y^{})
+ \sqrt{\frac{2}{3}}\cos \theta_0\frac{F_0}{F_\pi}(\bar X^{} -2\, \bar Y^{})\right)^2\,,
\end{align}
where the mixing is described in terms of octet and singlet decay constants $F_8$, $F_0$ and two mixing angles $\theta_8$, $\theta_0$
\cite{Kaiser:1998ds,Kaiser:2000gs,Feldmann:1998vh,Feldmann:1999uf}. Numerical input values are collected in Table~\ref{tab:input}.
The functions $\bar X$ and $\bar Y$ are given by
\begin{align}
\label{eq:Xbar}
&\bar X^{}=X_0(y_4)+\sum\limits_{i=d,s,b,b^\prime}|V_{ui}|^2F^{\nu\bar\nu}(y_4,x_i)\,,\\
\label{eq:Ybar-tau-mu}
&\bar Y^{}=Y_0(y_4)+\sum\limits_{i=u,c,t,t^\prime}|V_{id}|^2F^{\mu\bar\mu}(y_4,x_i)\,.
\end{align}
Here $x_i=m_{u_i}^2/M_W^2,\,m_{d_i}^2/M_W^2$ are functions of the quark masses,
 and $V_{ij}$ are the elements of the $V_\text{4G}$ mixing matrix in the quark sector.
The functions $F^{\mu\bar\mu}$ and $F^{\nu\bar\nu}$ are known from our analysis of rare $K$ and $B$ decays in \cite{Buras:2010pi}. 
We recall their explicit expressions in Appendix~\ref{app:functions}.

\boldmath
\subsection{$\mu-e$ Conversion in Nuclei\label{sec:mu-e-conversion}}
\unboldmath
Here, as in \cite{Blanke:2007db}, we use the general formula (58) of \cite{Hisano:1995cp} to find the approximate conversion rate
\begin{eqnarray}\label{eq:mueconv}
\Gamma(\mu\text{X} \to e\text{X})&=&\frac{G_F^2}{8\pi^4}\alpha^5\frac{Z_\text{eff}^4}{Z}|F(q^2)|^2m_\mu^5 \left|\chi_4^{(\mu e)}\right|^2\\
&&\times \left[ Z\left(4\bar Z^{}+H(y_4) \right)-(2Z+N)\frac{\bar X^{}}{\sin^2\theta_W}+(Z+2N)\frac{\bar Y^{}}{\sin^2\theta_W}  \right]^2\,,\qquad\nn
\end{eqnarray}
where $\bar X^{}$ and $\bar Y^{}$ are given in (\ref{eq:Xbar}) and (\ref{eq:Ybar-tau-mu}), and $H(x)$ and $\bar Z^{}$ are given in (\ref{eq:H(x)}) and (\ref{eq:Zbar}). $Z$ and $N$ denote the proton and neutron number of the nucleus. $Z_\text{eff}$ has been determined in \cite{Sens:1959,Ford:1962wi,Pla:1971be,Chiang:1993xz} and $F(q^2)$ is the nucleon form factor. For $\text{X}={}^{48}_{22}\text{Ti}$, one has $Z_\text{eff}=17.6$ and $F(q^2\simeq-m_\mu^2)\simeq 0.54$ \cite{Bernabeu:1993ta}.

The $\mu-e$ conversion rate ${\rm R}(\mu\text{X}\to e\text{X})$ is then given by
\begin{equation}
{\rm R}(\mu\text{X}\to e\text{X})=\frac{\Gamma(\mu\text{X} \to e\text{X})}{\Gamma^\text{X}_\text{capture}}\,,
\end{equation}
with $\Gamma^\text{X}_\text{capture}$ being the $\mu$ capture rate of the element X. For titanium the experimental value is given by \cite{Suzuki:1987jf}
\begin{equation}\label{eq:mu-e-conv-exp}
\Gamma^\text{Ti}_\text{capture}=
(2.590\pm0.012)\cdot10^6\,\text{s}^{-1}\,. 
\end{equation}

In our numerical analysis of Section \ref{sec:numerics} we will restrict ourselves to $\mu-e$ conversion in $^{48}_{22}\text{Ti}$, {for which the most stringent experimental upper bound exists and where the approximations entering \eqref{eq:mueconv}  work very well. For details, we refer the reader to the discussion presented in \cite{Hisano:1995cp,Kitano:2002mt,Bernabeu:1993ta}.}

\boldmath
\subsection{$K_{L,S} \to \mu e$ and $K_{L,S} \to \pi^0 \mu e$ \label{sec:K-mu-e}}
\unboldmath
In the SM3 the decay $K_L \to \mu e$ can proceed through box diagrams in the case of non-degenerate neutrino masses, but similarly to $\mu\to e \gamma$ its rate is too small to be measured.

\begin{figure}
\center{
\includegraphics[scale=.8]{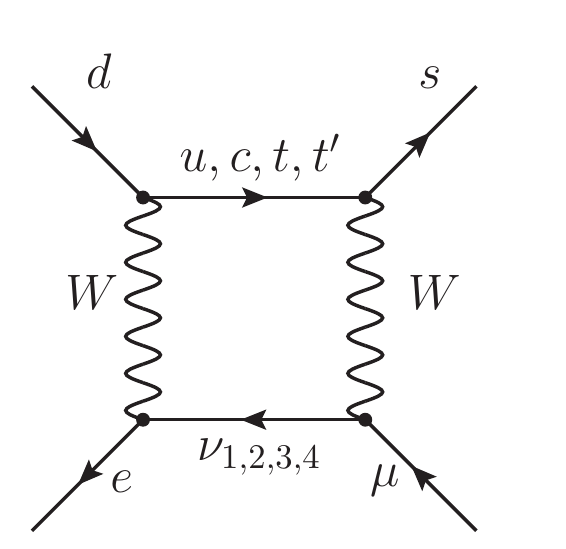}
}
\caption{Diagrams contributing to the $K_L \to \mu e$ decay in the SM4.\label{diag:K-to-mu-e}}
\end{figure}

{Also in the SM4, $K_L \to \mu e$ proceeds through box diagrams as shown in Fig.~\ref{diag:K-to-mu-e}, but due to the large mass of the 4G neutrino this contribution now becomes relevant.}
The effective Hamiltonian corresponding to these diagrams is given by
\be 
{\mathcal H}_{\rm eff} = \frac{G_F^2}{8 \pi^2} M_W^2 \sum\limits_{i,j} \lambda_i^{(K)} \chi_j^{(\mu e)} P(x_i,y_j) (\bar s d)_{V-A} (\bar e \mu)_{V-A}\,,
\ee
where
\be
\lambda_i^{(K)} = V_{is}^\ast V_{id}\,,
\ee
and the function $P(x_i,y_j)$ can be found in Appendix~\ref{app:functions}.

Using the unitarity of the 4G leptonic mixing matrix and the fact that $y_j\approx 0$ for $j\neq 4$, this expression simplifies to
\begin{equation}\label{eq:Heff-KL-mue}
{\mathcal H}_{\rm eff} = \frac{G_F^2}{8 \pi^2} M_W^2 \chi_4^{(\mu e)}\sum\limits_{i} \lambda_i^{(K)}  \left[P(x_i,y_4)-P(x_i,0)\right] (\bar s d)_{V-A} (\bar e \mu)_{V-A}\,.
\end{equation}

The evaluation of $\Br(K_L\to \mu e)$ by means of ${\mathcal H}_{\rm eff}$ in (\ref{eq:Heff-KL-mue}) proceeds along the derivation presented in Section~7 of \cite{Blanke:2007db}. We accordingly find
\begin{eqnarray} \label{eq:BrKLmue}
\Br(K_L\to \mu e) &=&  \frac{G_F^2}{8 \pi^4} M_W^4 \Br(K^+\to \mu^+ \nu) \frac{\tau(K_L)}{\tau(K^+)} \frac{1}{|V_{us}|^2} \left|\chi_4^{(\mu e)}\right|^2\nonumber \\
&& \times \left(\sum_{i=c,t,t'}\RE(\lambda_i^{(K)}) R(x_i,0,0,y_4) \right)^2\,,
\end{eqnarray}
where the function $R$ is defined as follows
\be
R(x_i,x_j,y_k,y_l) = P(x_i,y_l) + P(x_j,y_k) - P(x_j,y_l) - P(x_i,y_k)\,,
\ee
and
\cite{Amsler:2008zzb,Antonelli:2010yf,Antonelli:2009ws}
\be
\Br(K^+\to\mu^+\nu)=(63.44 \pm 0.14)\%\,,\quad
\frac{\tau(K_L)}{\tau(K^+)}=4.117 \pm 0.019\,,\quad |V_{us}|=0.225 \pm 0.001\,.
\ee
{Note that in general $|V_{us}|$ from a SM3 fit of semileptonic $K$ decays is not longer valid in the SM4. However since a
reanalysis of this fit in the context of the SM4 is clearly beyond the scope of the work, we use the above value for simplicity.}
Experimentally we have \cite{Ambrose:1998us,:2007yz}
\be
{\rm Br}(K_L\to \mu e) \equiv {\rm Br} (K_L \to \mu^+ e^-) + {\rm Br} (K_L \to \mu^- e^+) < 4.7\cdot 10^{-12}\,.
\ee

Similarly, following \cite{Blanke:2007db} we find
\bea
\Br(K_L\to\pi^0\mu e)&\equiv&\Br(K_L\to\pi^0\mu^+ e^-)+\Br(K_L\to\pi^0\mu^- e^+)\nn\\
&=& \frac{G_F^2 M_{W}^4}{8\pi^4} \Br (K^+\to\pi^0\mu^+\nu)\frac{\tau(K_L)}{\tau(K^+)}\frac{1}{|V_{us}|^2}\left|\chi_4^{(\mu e)}\right|^2\nn\\
&&\times \left( \sum_{i=c,t,t'}\text{Im}(\lambda_i^{(K)}) R(x_i,0,0,y_4) \right)^2\,,\label{eq:BrKLpi0mue}
\eea
where \cite{Amsler:2008zzb}
\be
\Br(K^+\to\pi^0\mu^+\nu)=(3.32\pm0.06)\%\,.
\ee

\boldmath
\subsection{Lepton-Flavour Violating $B$ Decays \label{sec:B-mu-e}}
\unboldmath

A detailed study of the decays $B_{d,s} \to \mu e$, $B_{d,s} \to \tau e$ and $B_{d,s} \to \tau \mu$ in the LHT model has been presented in Section 8 of \cite{Blanke:2007db}. 
{ The formulae given therein can easily be adapted to the SM4 case and are summarised by the following two equations:
\begin{eqnarray}
 \Br(B_d\to \ell_1 \ell_2) &=& \frac{G_F^2 M_W^4}{16 \pi^4 |V_{ub}|^2} \frac{\tau(B_d)}{\tau(B^+)} \Br(B^+\to \ell_1^+ \nu_{\ell_1}) \left|\chi_4^{(\ell_1 \ell_2)}\right|^2 \nn \\
&&\times \left|\sum_{i=c,t,t'} \lambda_i^{(d)} R(x_i,0,0,y_4)\right|^2 \\
\Br(B_s\to \ell_1 \ell_2) &=& \frac{G_F^2 M_W^4}{16 \pi^4 |V_{ub}|^2} \frac{\tau(B_s)}{\tau(B^+)} \frac{M_{B_s}}{M_{B_d}} \frac{F_{B_s}^2}{F_{B_d}^2} \Br(B^+\to \ell_1^+ \nu_{\ell_1}) \left|\chi_4^{(\ell_1 \ell_2)}\right|^2 \nn \\
&&\times \left|\sum_{i=c,t,t'} \lambda_i^{(s)} R(x_i,0,0,y_4)\right|^2\,,
\end{eqnarray}
}
where $\ell_1$ and $\ell_2$ denote the two leptons in the final state with $m_{\ell_1} > m_{\ell_2}$. {For our numerical analysis we used the SM predictions for $\Br(B^+\to \mu^+ \nu_{\mu})$ and $\Br(B^+\to \tau^+ \nu_{\tau})$ \cite{Bona:2009cj,Altmannshofer:2009ne}
\begin{eqnarray}
 \Br(B^+\to \mu^+ \nu_{\mu}) &= & \left( 3.8\pm 1.1 \right)\cdot 10^{-7}\,,\\
 \Br(B^+\to \tau^+ \nu_{\tau}) &= & \left( 0.8\pm 0.12 \right)\cdot 10^{-4}\,.
\end{eqnarray}
This is necessary, because currently only $\Br(B^+\to \tau^+ \nu_{\tau}) =\left( 1.67 \pm 0.39 \right)\cdot 10^{-4}$  is measured \cite{:2008ch,:2008gx,Barberio:2008fa}. The SM3 value is on the lower side 
of this measurement but still consistent within errors.}

\boldmath
\subsection{Anomalous Magnetic Moment of the Muon}
\unboldmath\label{sec:g-2}
The one loop contribution to $a_\mu\equiv(g-2)_\mu/2 = \left(a_\mu\right)_\text{SM3} + \left(a_\mu\right)_\text{SM4}$ in the SM4 can be obtained in analogy to our derivation in \cite{Blanke:2007db}. Only the triangle diagram with a $W$ boson and two heavy neutrinos 
running in the loop is relevant here. Adapting the corresponding expression (11.11) in \cite{Blanke:2007db} to the SM4 by dividing by a factor $v^2/(4f^2)$, removing the sum over 
different flavours and adjusting the matrix elements $|V_{H\ell}^{i\mu}|^2\to |U_{\mu4}|^2$ we obtain 
\begin{eqnarray}
\left(a_\mu\right)_\text{SM4}=-\frac{\sqrt{2}G_F}{8\pi^2}m_\mu^2\left|U_{\mu 4}\right|^2\left(L_2(y_4) - L_2(0)\right)\,,
\end{eqnarray}
where the function $L_2(x)$ is given in Appendix \ref{app:functions}.
Two comments are in order at this point:
\begin{enumerate}
 \item Since $L_2(y_4) - L_2(0)>0$, the SM4 contribution tends to decrease $a_\mu$ and thus pushes it even further away from the experimental value.
 \item It turns out that after imposing the constraints from lepton universality and radiative decays, the SM4 contribution to $a_\mu$ becomes negligible compared to the theoretical uncertainties.
\end{enumerate}

\section{Numerical Analysis\label{sec:numerics}}
\subsection{Preliminaries}

\begin{table}[ht]
\begin{center}
\begin{tabular}{|l|l|}\hline
$m_e=0.5110\mev$ & $\tau(B_d)/\tau(B^+)=0.934(7)$\\
$m_\mu=105.66\mev$ & $\tau(B_s)=1.425(41)\,\text{ps}$ \\ 
$m_\tau=1.77684(17)\gev$ & $\tau(B^+)=1.638(11)\,\text{ps}$ \\
$\tau_\tau = 290.6(1.0)\cdot 10^{-3}\,\text{ps}$ & $M_{B_d}=5.2794(5)\gev$\\
$M_W=80.425(38)\gev$ & $M_{B_s}=5.3675(18)\gev$ \quad\\
$\alpha=1/137$ & $|V_{ub}|=3.68(14)\cdot10^{-3}$ \hfill\cite{Amsler:2008zzb} \\\cline{2-2}
$G_F=1.16637(1)\cdot 10^{-5} \gev^{-2}$\quad & $F_8/F_\pi=1.28$ \hfill(ChPT)\\
$\sin^2\theta_W = 0.23122(15)$ \qquad\cite{Amsler:2008zzb} & $F_0/F_\pi=1.18(4)$ \\ \cline{1-1}
$F_{B_d}=192.8(9.9)\mev$ & $\theta_8=-22.2(1.8)^\circ$ \\
$F_{B_s}=238.8(9.5)\mev$\hfill \cite{Laiho:2009eu} & $\theta_0=-8.7(2.1)^\circ$\hfill\cite{Escribano:2005qq}\\
& $F_\pi=130 \pm 5~\mev$ \hfill\cite{Amsler:2008zzb}\\
\hline
\end{tabular}
\end{center}
\caption{Values of the experimental and theoretical quantities used as input parameters.\label{tab:input}}
\end{table}

The great simplicity of the analysis of LFV within the SM4 when compared to NP scenarios such as the general MSSM, the LHT and RS models is the paucity of free parameters. 
The analysis also simplifies compared to the quark sector since the contributions of SM3 leptons in loops can be neglected, except when they are relevant in the context of the GIM mechanism.

We note that $\ell_i\to\ell_j\gamma$ and the decays to three leptons are fully governed by the quantities
\begin{equation}
\chi_4^{(\mu e)}\,,\qquad \chi_4^{(\tau e)}\,,\qquad \chi_4^{(\tau \mu)}\,,
\qquad |U_{e4}|\,,\qquad |U_{\mu 4}|\,,
\end{equation}
and calculable functions of the neutrino mass $m_{\nu_4}$ which is bounded by direct measurements \cite{Amsler:2008zzb},
\begin{equation}
m_{\nu_4}\geq 90.3 \gev\ \qquad (95\%\ \text{C.L.})\,.
\end{equation}
Therefore strong correlations between the $\ell_i\to\ell_j\gamma$ and $\ell_i\to3\ell_k$ decays are to be expected. While this expectation will be confirmed in the course of our numerical analysis,
we will see that the possible ranges for various observables entering these correlations will still be rather large.

Semileptonic decays and $\mu-e$ conversion in nuclei involve also parameters in the quark sector that enter through box diagram contributions to the 
functions $\bar X^{}$ and $\bar Y^{}$ in (\ref{eq:Ybar-tau-mu}) and (\ref{eq:Xbar}). These contributions are however constrained through our analysis of the quark sector in the SM4.
In our analysis we also take into account constraints present outside the LFV sector, in particular those from \cite{Lacker:2010zz}.

For our numerical analysis of processes involving quarks we used the points of our previous analysis \cite{Buras:2010pi}. Our parameter points were generated using 
uniform random numbers, and we explicitly do not assign any statistical meaning to the point densities. We included the effect of a modified Fermi constant $G_F$ due to the breaking of three-generation 
lepton-universality \cite{Lacker:2010zz} and included the decays $\tau\to \mu\nu_\mu\nu_\tau$ and $\tau \to e\nu_e\nu_\tau$ to constrain the parameters. Contrary to \cite{Lacker:2010zz}
we do not find a significant effect of the $K_{3\ell}$ decays, but this is due to our much more conservative error treatment. On this note we want to reemphasise the need
for a consistent fit of the EWP data, CKM matrix elements from semileptonic decays, $G_F$ and similar well known inputs \cite{Lacker:2010zz,Erler:2010sk,Chanowitz:2009mz,Bobrowski:2009ng,Eberhardt:2010bm,Chanowitz:2010bm},
in the context of the SM4.

\boldmath
\subsection{$\mu^-\to e^-\gamma$, $\mu^-\to e^-e^+e^-$ and $\mu-e$ Conversion}
\begin{figure}
\begin{center}
 \includegraphics[width=.7\textwidth]{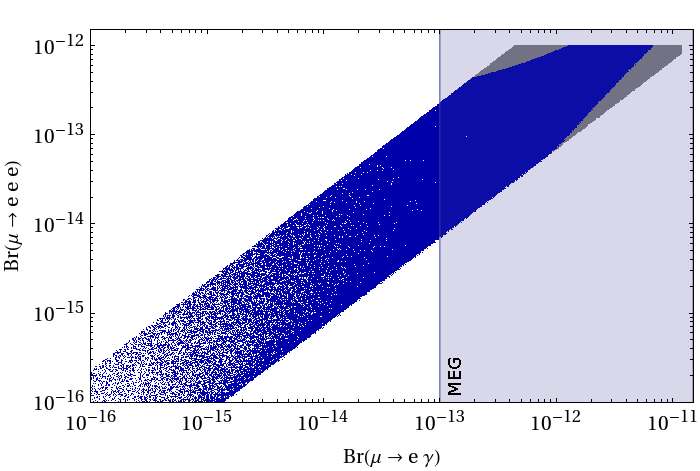}
\end{center}
\caption{Correlation between $\Br(\mu\to e\gamma)$ and $\Br(\mu^-\to e^-e^+e^-)$. {Points that agree with the currently measured $\mu-e$ conversion rate in $^{48}_{22}\text{Ti}$ (\ref{eq:mu-e-conv-exp}) are shown in blue, while gray points violate this bound. The shaded area indicates the projected experimental bound on $\Br(\mu\to e\gamma)$ from the MEG experiment at PSI.}  \label{fig:meg-meee}}
\end{figure}

\unboldmath
In Fig.~\ref{fig:meg-meee} we show the correlation between $\mu\to e\gamma$ and $\mu^-\to e^-e^+e^-$ together with the experimental bounds on these decays. We observe:
\begin{itemize}
 \item 
 Both branching ratios can easily reach the present experimental bounds 
 in a correlated manner.
 \item 
 However, for a fixed value of either branching ratio, the other one
 can still vary over one order of magnitude.
 \end{itemize}

\begin{figure}
\begin{center}
 \includegraphics[width=0.7 \textwidth]{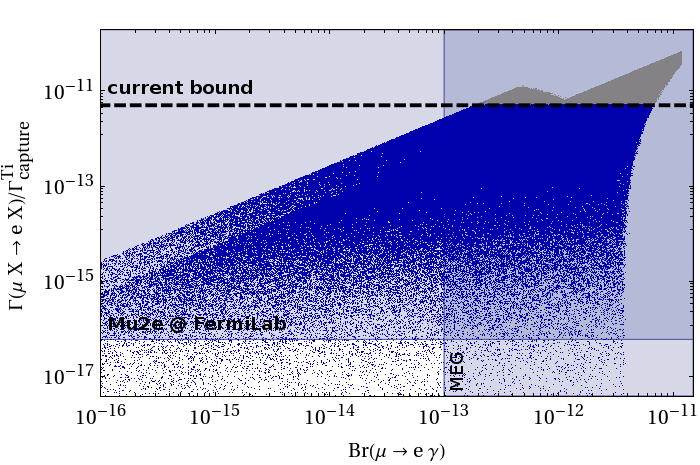}
\end{center}
\caption{Correlation between $\Br(\mu\to e\gamma)$ and ${\rm R}(\mu\text{Ti}\to e\text{Ti})$. {The shaded areas indicate the expected future experimental bounds on both observables.}\label{fig:meg-mue-conv}}
\end{figure}

Next in Fig.~\ref{fig:meg-mue-conv} we show the correlation between the $\mu-e$ conversion rate in $^{48}_{22}\text{Ti}$ and $\Br(\mu\to e \gamma)$, after imposing the existing constraints on 
\mbox{${\rm Br}(\mu\to e\gamma)$} and \mbox{${\rm Br}(\mu^-\to e^-e^+e^-)$}. 
We observe that this correlation is weaker than the one in Fig.~\ref{fig:meg-meee} as now also quark parameters enter the game. Still, for a given 
$\Br(\mu\to e\gamma)$ a sharp upper bound on the $\mu\to e$ conversion rate 
is identified. Furthermore, we find that the $\mu-e$ conversion rate in titanium is generally larger than the current experimental bound,
but the bounds on both branching ratios can be simultaneously satisfied. 
Yet it is evident from this plot that lowering the upper bounds on both 
observables in the future will significantly reduce the allowed regions of the leptonic parameter space in the SM4.

\begin{figure}
\begin{center}
 \includegraphics[width=0.7 \textwidth]{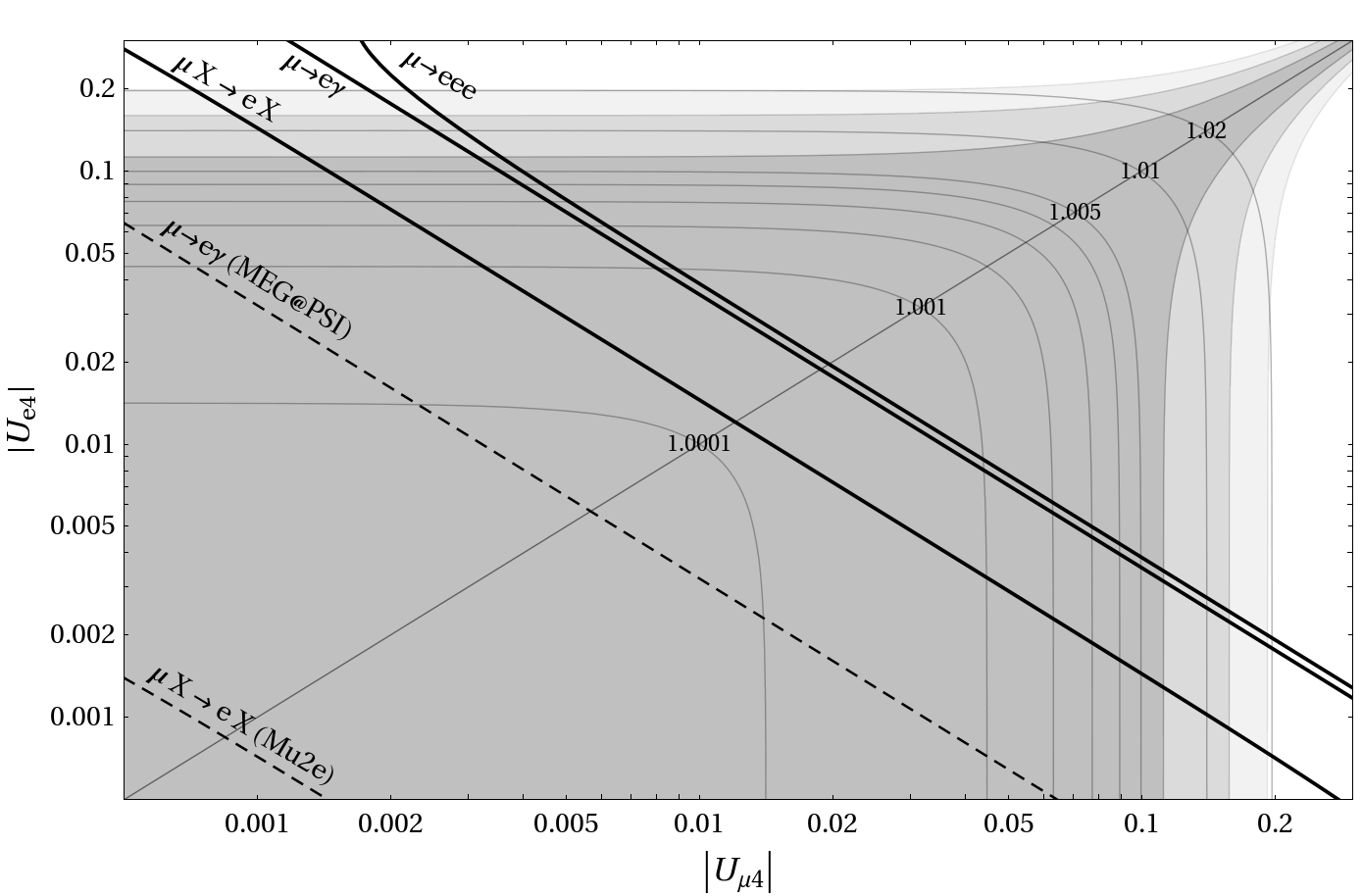}
\end{center}
\caption{Constraints on the allowed range of $|U_{e4}|$ and $|U_{\mu 4}|$ resulting from lepton universality ($1\sigma$/ $2\sigma$/$3\sigma$:  dark gray/gray/light gray area, respectively) 
and the current experimental bounds on $\mu\to eee$, $\mu\to e\gamma$, and $\mu-e$ conversion (thick black lines). The contour lines indicate the ratio $G_F^{SM4}/G_F^{SM3}$, where $G_F^{SM4}$
 is the value of the Fermi constant extracted from muon lifetime measurement assuming 4 generations, and $G_F^{SM3}$ is the usual SM3 Fermi constant (s.~\cite{Lacker:2010zz}).\label{fig:Umu4Ue4}} 
\end{figure}

As pointed out by \cite{Lacker:2010zz}, the combination of results from leptonic $\tau$ decays and radiative $\mu$ decays efficiently constrains the involved PMNS 
parameters $|U_{e4}|$ and $|U_{\mu 4}|$. We show these bounds in Fig.~\ref{fig:Umu4Ue4}, adding the constraint from $\mu-e$ conversion which turns out to be the most stringent one.


\boldmath
\subsection{The Decays $\tau\to\mu\gamma$ and $\tau\to e\gamma$}
\unboldmath
\begin{figure}
\begin{center}
 \includegraphics[width=0.7 \textwidth]{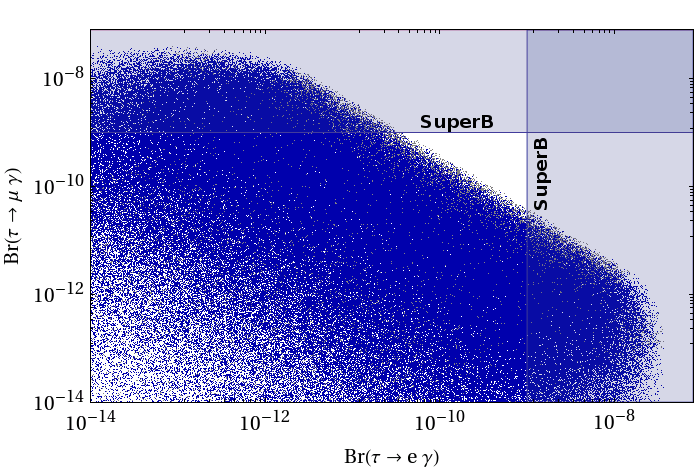}
\end{center}
\caption{Correlation between $\Br(\tau\to \mu\gamma)$ and $\Br(\tau\to e\gamma)$.\label{fig:tmg-teg}}
\end{figure}

In Fig.~\ref{fig:tmg-teg} we show the correlation between $\Br(\tau\to\mu\gamma)$ and $\Br(\tau\to e\gamma)$, imposing the experimental bounds on $\mu\to e\gamma$ and $\mu^-\to e^-e^+e^-$.
We observe that they both can be individually as high as few times $10^{-8}$ and thus in the ball park of present experimental upper bounds. 
The maximal values however cannot be reached simultaneously, due to the $\mu\to e\gamma$ constraint.
Thus finding both branching ratios at the $10^{-8}-10^{-9}$ level will basically eliminate the SM4 scenario.

\FloatBarrier

\boldmath
\subsection{The Decays $\tau\to\mu\pi,\,\mu\eta,\,\mu\eta^\prime$ and $\tau\to\mu\gamma$}
\begin{figure}[ht]
\begin{center}
  \includegraphics[width=0.7 \textwidth]{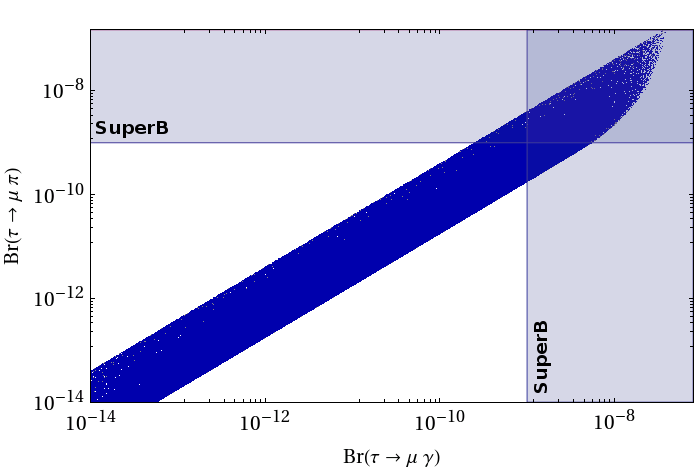}
\end{center}
\caption{$\Br(\tau\to\mu\pi)$ as a function of $\Br(\tau\to\mu\gamma)$\label{fig:tmg-tmp}} 
\end{figure}

\unboldmath
In Fig.~\ref{fig:tmg-tmp} we show $\Br(\tau\to\mu\pi)$ a function of $\Br(\tau\to\mu\gamma)$, imposing the constraints from $\mu\to e\gamma$ 
and $\mu^-\to e^-e^+e^-$. We find that $\Br(\tau\to\mu\pi)$ can reach values as high as the present experimental bounds from Belle and BaBar, which is in the ball park of $10^{-8}$. {It is evident from (\ref{eq:tmpi})-(\ref{eq:tmetaprime}) that $\Br(\tau\to\mu\eta^\prime)$ and $\Br(\tau\to\mu\eta)$ are strongly correlated with $\Br(\tau\to\mu\pi)$, so we choose not to show the respective plots for these processes.}

Completely analogous correlations can be found also for the corresponding decays $\tau\to e\pi,e\eta,e\eta'$ and $\tau\to e\gamma$. Indeed, 
this symmetry between the $\tau\to\mu$ and $\tau\to e$ systems {turns out to be a general feature of the SM4}, that can be found in all decays considered in the present paper.
We will return to this issue in Section \ref{sec:distinction}.

An immediate consequence of these correlations is that the observation 
of a large  $\tau\to\mu\gamma$ rate will immediately imply a large
 $\tau\to\mu\pi$ rate and vice versa. Still, for a fixed value of either branching ratio the second one can vary by almost an order of magnitude. Analogous statements apply to $\tau\to\mu(e)\eta$ and $\tau\to\mu(e)\eta^\prime$.

\boldmath
\subsection{$K_L \to \mu e$ and $K_L\to \pi^0 \mu e$}
\unboldmath

\begin{figure}[ht]
\begin{center}
\includegraphics[width=0.7\textwidth]{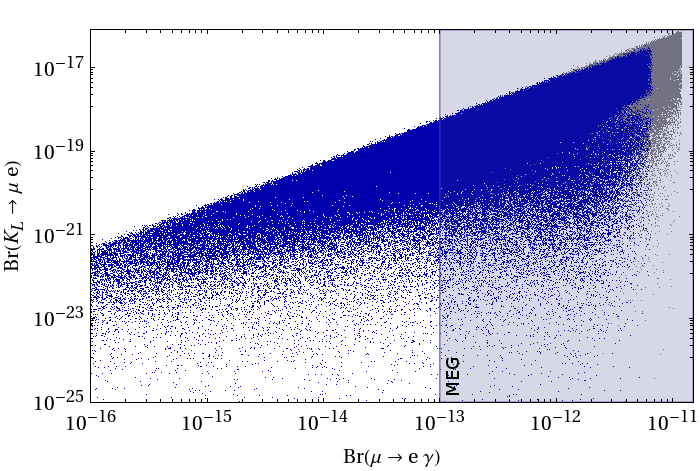}
\end{center}
\caption{$\Br(K_L \to\mu e)$ as a function of $\Br(\mu\to e\gamma)$.
{Colour coding defined in Fig.~\ref{fig:meg-meee}.}
\label{fig:meg-KLme}} 
\end{figure}

\begin{figure}[ht]
\begin{center}
\includegraphics[width=0.7\textwidth]{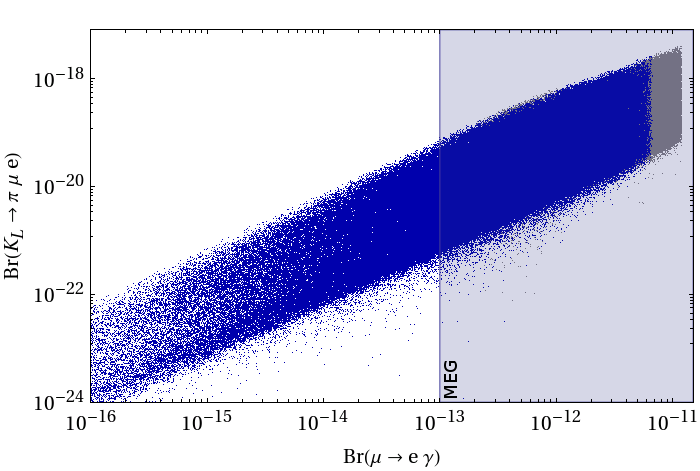}
\end{center}
\caption{$\Br(K_L \to\pi^0 \mu e)$ as a function of $\Br(\mu\to e\gamma)$.
{Colour coding defined in Fig.~\ref{fig:meg-meee}.}
\label{fig:meg-KLpme}} 
\end{figure}

In  Figs.~\ref{fig:meg-KLme} and \ref{fig:meg-KLpme} we show the results 
for $\Br(K_L \to \mu e)$ and $\Br(K_L\to \pi^0 \mu e)$ as functions of 
$\Br(\mu\to e\gamma)$. Again strong correlations between these branching 
ratios are observed but the maximal values for $\Br(K_L \to \mu e)$ and 
$\Br(K_L\to \pi^0 \mu e)$ are by several orders of magnitude below the 
present experimental bounds.

\FloatBarrier

\subsection{Upper Bounds}
In Table \ref{tab:upper-bounds} we show the maximal values obtainable in the SM4 for all branching ratios considered in the present paper, together with the corresponding experimental bounds. 
We observe:
\begin{itemize}
 \item 
 The branching ratios for eleven of decays in this table can still come close
 to the respective experimental bounds and as we have seen in the previous plots they are correlated with each other. 
 \item 
 The remaining branching ratios are by several orders of magnitude below
the present experimental bounds and if the SM4 is the whole story, these 
decays will not be seen in the foreseeable future.
\item
Comparing to the results obtained in the LHT model for a NP scale $f=1 \tev$ \cite{Blanke:2007db,Blanke:2009am}, the SM4 allows for much larger branching ratios but the difference is much smaller for $f=500\gev$.
\end{itemize}
We have also investigated the effect of additionally imposing  \mbox{${\rm R}(\mu\text{Ti}\to e\text{Ti})<5\cdot10^{-12}$} as a constraint, which we have chosen slightly above the experimental
 value $4.3\cdot 10^{-12}$ in order to account for the involved theoretical uncertainties. We find that all maximal values collected in Table \ref{tab:upper-bounds} depend only weakly on that constraint.
\setlength{\LTcapwidth}{0.85\textwidth}
\begin{center}
\begin{longtable}{|c|c|c|}
\hline
decay & maximal value& exp.~upper bound \\\hline\hline
\endfirsthead

\hline
decay & maximal value& exp.~upper bound \\\hline
&\vdots&
\endhead
&\vdots&\\\hline
\endfoot
\endlastfoot
$\mu\to e\gamma$ & $1.2\cdot 10^{-11}$ ($6.8\cdot10^{-12}$)  & $1.2\cdot 10^{-11}$ \cite{Brooks:1999pu} \\
$\mu^-\to e^-e^+e^-$ & ~$1.0\cdot 10^{-12}$ ($1\cdot 10^{-12}$)~ & $1.0\cdot 10^{-12}$ \cite{Bellgardt:1987du} \\
${\rm R}(\mu\text{Ti}\to e\text{Ti})$  & $6.6\cdot 10^{-11}$ ($5\cdot10^{-12}$)& $4.3\cdot10^{-12}$ \cite{Dohmen:1993mp}  \\\hline
$\tau\to e\gamma$ & $3.9\cdot 10^{-8}$ ($3.9\cdot 10^{-8}$)  & ${3.3\cdot10^{-8}}$ \cite{:2009tk} \\
$\tau\to \mu\gamma$ & $3.9\cdot 10^{-8}$ ($3.9\cdot 10^{-8}$) &${4.4\cdot10^{-8}}$ \cite{:2009tk}\\
$\tau^-\to e^-e^+e^-$ & $7.5\cdot10^{-8}$ ($7.5\cdot 10^{-8}$)& $2.7\cdot10^{-8}$ \cite{Hayasaka:2010np}\\
$\tau^-\to \mu^-\mu^+\mu^-$ & $7.4\cdot10^{-8}$ ($7.1\cdot 10^{-8}$)& $2.1\cdot10^{-8}$ \cite{Hayasaka:2010np} \\
$\tau^-\to e^-\mu^+\mu^-$ & $5\cdot10^{-8}$ ($5\cdot 10^{-8}$)  & $2.7\cdot10^{-8}$ \cite{Hayasaka:2010np}\\
$\tau^-\to \mu^-e^+e^-$ & $5\cdot10^{-8}$ ($5\cdot 10^{-8}$)&$1.8\cdot10^{-8}$ \cite{Hayasaka:2010np} \\
$\tau^-\to \mu^-e^+\mu^-$ & $4.7\cdot10^{-17}$ ($4.7\cdot10^{-17}$) & $1.7 \cdot10^{-8}$ \cite{Hayasaka:2010np}\\
$\tau^-\to e^-\mu^+e^-$ & $4.9\cdot10^{-17}$ ($4.9\cdot10^{-17}$) & $1.5 \cdot10^{-8}$ \cite{Hayasaka:2010np} \\
$\tau\to\mu\pi$ & $1.4\cdot10^{-7} $ ($1.4\cdot10^{-7} $) & ${5.8\cdot10^{-8}}$ \cite{Banerjee:2007rj}\\
$\tau\to\mu\eta$ & $2.5\cdot10^{-8}$ $(2.5\cdot10^{-8})$  &  ${5.1\cdot 10^{-8}}$ \cite{Banerjee:2007rj}\\
$\tau\to \mu\eta'$ & $2.9\cdot10^{-10}$ $(2.9\cdot10^{-8})$ & ${5.3\cdot 10^{-8}}$ \cite{Banerjee:2007rj}\\
$K_L\to\mu e$ & $7.7\cdot 10^{-17}$ ($3.3\cdot10^{-17}$) & $4.7\cdot10^{-12}$ \cite{Ambrose:1998us}\\
$K_L\to\pi^0\mu e$ & $3.5\cdot 10^{-18}$ ($2.1\cdot10^{-18}$)  & $6.2\cdot10^{-9}$ \cite{Arisaka:1998uf}\\
$B_d\to\mu e$ & $2.4\cdot10^{-18}$ ($1.3\cdot10^{-18}$)  & $9.2\cdot10^{-8}$ \cite{Amsler:2008zzb}\\
$B_s\to\mu e$ & $7.2\cdot 10^{-17}$ ($4.0\cdot10^{-17}$) & $6.1\cdot10^{-6}$ \cite{Abe:1998bc}\\
$B_d\to\tau e$ & $1.4\cdot 10^{-11}$  ($1.4\cdot10^{-11}$) & $2.8\cdot10^{-5}$ \cite{Amsler:2008zzb}\\
$B_s\to\tau e$ & $5.4\cdot10^{-10}$ ($5.4\cdot10^{-10}$)& ---\\
$B_d\to\tau\mu$ & $1.4\cdot10^{-11}$ ($1.4\cdot10^{-11}$)  & $2.2\cdot10^{-5}$ \cite{Amsler:2008zzb} \\
$B_s\to\tau\mu$ & $5.4\cdot10^{-10}$ ($5.4\cdot10^{-10}$) & ---\\\hline
\multicolumn{3}{l}{}\\
\caption{Maximal values for LFV decay branching ratios in the SM4, after imposing the constraints on 
${\rm Br}(\mu\to e\gamma)$ and ${\rm Br}(\mu^-\to e^-e^+e^-)$. The numbers given in brackets are obtained after imposing the additional constraint \mbox{${\rm R}(\mu\text{Ti}\to e\text{Ti})<5\cdot10^{-12}$}. 
The current experimental upper bounds are also given.}
\label{tab:upper-bounds}
\end{longtable}

\end{center}


 This finding justifies that we did not take into account this bound in our numerical analysis so far, as it has only a minor impact on the discussed observables.
We would like to stress that the {maximal values} in Table \ref{tab:upper-bounds} should only be considered as rough upper bounds. They have been obtained from scattering over the
 allowed parameter space of the model. In particular, no confidence level can be assigned to them. The same applies to the ranges given in Table \ref{tab:ratios} for the SM4 and the LHT model.

\FloatBarrier
\subsection{Patterns of Correlations and Comparison with the MSSM and the LHT\label{sec:distinction}}
In \cite{Blanke:2007db,Blanke:2009am} a number of correlations have been identified that allow to distinguish the LHT model from the MSSM. These results are recalled in Table \ref{tab:ratios}.
 In the last column of this table we also show the results obtained in the SM4. We observe:
\begin{itemize}
 \item 
For most of the ratios considered here the values found in the SM4 are significantly 
larger than in the LHT and by one to two orders of magnitude larger than 
in the MSSM. 
 \item 
In the case of $\mu\to e$ conversion the predictions of the SM4 and the 
LHT model are very uncertain but finding said ratio to be of order 
one would favour the SM4 and the LHT model over the MSSM.
\item
Similarly, in the case of several ratios considered in this table, finding 
them to be of order one will choose the SM4 as a clear winner in this 
competition. 
\end{itemize}
\begin{table}[ht]
{\renewcommand{\arraystretch}{1.5}
\begin{center}
\begin{tabular}{|c|c|c|c|c|}
\hline
ratio & LHT  & MSSM (dipole) & MSSM (Higgs)&SM4 \\\hline\hline
$\frac{\Br(\mu^-\to e^-e^+e^-)}{\Br(\mu\to e\gamma)}$  & \hspace{.8cm} 0.02\dots1\hspace{.8cm}  & $\sim6\cdot10^{-3}$ &$\sim6\cdot10^{-3}$ & $0.06\dots 2.2$ \\
$\frac{\Br(\tau^-\to e^-e^+e^-)}{\Br(\tau\to e\gamma)}$   & 0.04\dots0.4     &$\sim1\cdot10^{-2}$ & ${\sim1\cdot10^{-2}}$& $0.07\dots 2.2$\\
$\frac{\Br(\tau^-\to \mu^-\mu^+\mu^-)}{\Br(\tau\to \mu\gamma)}$  &0.04\dots0.4     &$\sim2\cdot10^{-3}$ & $0.06\dots0.1$& $0.06\dots 2.2$ \\\hline
$\frac{\Br(\tau^-\to e^-\mu^+\mu^-)}{\Br(\tau\to e\gamma)}$  & 0.04\dots0.3     &$\sim2\cdot10^{-3}$ & $0.02\dots0.04$& $0.03\dots 1.3$ \\
$\frac{\Br(\tau^-\to \mu^-e^+e^-)}{\Br(\tau\to \mu\gamma)}$  & 0.04\dots0.3    &$\sim1\cdot10^{-2}$ & ${\sim1\cdot10^{-2}}$& $0.04\dots 1.4$\\
$\frac{\Br(\tau^-\to e^-e^+e^-)}{\Br(\tau^-\to e^-\mu^+\mu^-)}$     & 0.8\dots2   &$\sim5$ & 0.3\dots0.5& $1.5\dots 2.3$\\
$\frac{\Br(\tau^-\to \mu^-\mu^+\mu^-)}{\Br(\tau^-\to \mu^-e^+e^-)}$   & 0.7\dots1.6    &$\sim0.2$ & 5\dots10& $1.4 \dots 1.7$ \\\hline
$\frac{{\rm R}(\mu\text{Ti}\to e\text{Ti})}{\Br(\mu\to e\gamma)}$  & $10^{-3}\dots 10^2$     & $\sim 5\cdot 10^{-3}$ & $0.08\dots0.15$&  $10^{-12}\dots 26$\\\hline
\end{tabular}
\end{center}\renewcommand{\arraystretch}{1.0}
}
\caption{Comparison of various ratios of branching ratios in the LHT model \cite{Blanke:2009am}, the MSSM without \cite{Ellis:2002fe,Brignole:2004ah} and
 with significant Higgs contributions \cite{Paradisi:2005tk,Paradisi:2006jp} and the SM4 {calculated here}.\label{tab:ratios}}
\end{table}

\FloatBarrier

\section{Conclusions\label{sec:conclusions}}
In the present paper we have calculated  branching ratios 
for a large number of  charged lepton flavour violating decays 
in the Standard Model extended by a fourth generation of quarks and leptons, {assuming that neutrinos are Dirac particles} and taking all presently available
constraints into account. Our main messages from this analysis are the following:
\begin{itemize}
\item
The branching ratios for $\ell_i\to\ell_j\gamma$, $\tau\to\ell\pi$, $\tau\to\ell\eta^{(\prime)}$, $\mu^-\to e^-e^+e^-$, $\tau^-\to e^-e^+e^-$,
 $\tau^-\to \mu^-\mu^+\mu^-$, $\tau^-\to e^-\mu^+\mu^-$ and 
$\tau^-\to \mu^-e^+e^-$ can all still be as large as the present experimental 
upper bounds but not necessarily simultaneously.
\item 
The correlations between 
various branching ratios should allow to test this model. This should  be contrasted with the SM3 where all these branching ratios are unmeasurable.
\item
The rate for  $\mu-e$ conversion in nuclei can also reach the corresponding 
upper bound.
\item
The pattern of the LFV branching ratios in the SM4 differs significantly from the one encountered in the MSSM, allowing to distinguish these two models with the help of LFV processes in a transparent manner. 
{The same statement applies to the LHT, as can be clearly seen from Table~\ref{tab:ratios}.}
\item
The  branching ratios for $K_L \to \mu e$, $K_L \to \pi^0\mu e$, $B_{d,s} \to \mu e$, $B_{d,s} \to \tau e$ and $B_{d,s} \to \tau\mu $ turn out to be by several 
orders of magnitude smaller than the present experimental bounds.
\end{itemize}

\subsection*{Acknowledgements}
This research was partially supported by the Cluster of Excellence `Origin and Structure of the Universe', the Graduiertenkolleg 
GRK 1054 of DFG and by the German `Bundesministerium f{\"u}r Bildung und
Forschung' under contract 05H09WOE.

\clearpage

\begin{appendix}
\newsection{Relevant Functions\label{app:functions}}

In this appendix we collect the various functions entering the theoretical formulas
for the LFV decays discussed in Sec.~\ref{sec:compendium}.
\begin{eqnarray}
\label{C0}C_0(x_i)&=&\frac{x_i}{8}\left[\frac{x_i-6}{x_i-1}+\frac{3x_i+2}{(x_i-1)^2}\log x_i\right],\\
\label{D0}D_0(x_i)&=&-\frac{4}{9}\log
x_i+\frac{-19x_i^3+25x_i^2}{36(x_i-1)^3}+\frac{x_i^2(5x_i^2-2x_i-6)}{18(x_i-1)^4}\log x_i\,,\\
\label{E0} E_0(x_i)&=&-\frac{2}{3}\log x_i + \frac{x_i^2(15-16x_i+4x_i^2)}{6(x_i-1)^4}\log x_i
+ \frac{x_i(18-11x_i-x_i^2)}{12(1-x_i)^3}\,,\\
\label{Dp0}D'_0(x_i)&=&-\frac{(3x_i^3-2x_i^2)}{2(x_i-1)^4}\log x_i +
\frac{(8x_i^3+5x_i^2-7x_i)}{12(x_i-1)^3}\,,\\
\label{Ep0}E'_0(x_i)&=&\frac{3x_i^2}{2(x_i-1)^4}\log x_i +
\frac{(x_i^3-5x_i^2-2x_i)}{4(x_i-1)^3}\,.\\
\label{X0}X_0(x_i)&=&\frac{x_i}{8}\;\left[\frac{x_i+2}{x_i-1}
+\frac{3 x_i-6}{(x_i -1)^2}\; \log x_i\right]\,,\\
\label{Y0}Y_0(x_i)&=&\frac{x_i}{8}\; \left[\frac{x_i-4}{x_i-1}
+ \frac{3 x_i}{(x_i -1)^2} \log x_i\right],\\
\label{Z0}Z_0(x_i)&=&-\frac{1}{9}\log x_i+\frac{18 x_i^4-163x_i^3+259
 x_i^2-108x_i}{144(x_i-1)^3}\nn\\
&&+\frac{32x_i^4-38x_i^3-15x_i^2+18x_i}{72(x_i-1)^4}\log x_i\,.
\end{eqnarray}
For arbitrary arguments $x_i,x_j$, the function
$S_0(x_i,x_j)$ is given by \cite{Buras:1983ap}
\begin{eqnarray}
  S_0 (x_i,x_j) &=& x_i x_j\left(\frac{(4 - 8 x_j + x_j^2)\log x_j}{4 (x_j-1)^2 (x_j-x_i)} +(i\leftrightarrow j) - \frac{3}{4 (x_i-1) (x_j-1)}\right)\,.
\end{eqnarray}
In the limit of $\varepsilon \rightarrow 0$
in $S_0(x_i+\varepsilon,x_i-\varepsilon)$ one recovers the SM3 version of $S_0(x_i)$,
\begin{eqnarray}
  S_0 (x_i) &=& \frac{x_i}{4}\;\frac{-4 + 15x_i - (12 - 6 \log x_i )x_i^2 + x_i^3}{(x_i-1)^3}\,.
\end{eqnarray}
The functions entering the $K_{L,S}$ decays discussed in Sec.~\ref{sec:K-mu-e} read
\begin{eqnarray}
F^{\mu\bar\mu}(x_i,y_4)&=&B^{\mu\bar\mu}(x_i,0)+B^{\mu\bar\mu}(0,y_4)-B^{\mu\bar\mu}(0,0)-B^{\mu\bar\mu}(x_i,y_4)\nonumber\\&=&-S_0(x_i,y_4)\,,\\
F^{\nu\bar\nu}(x_i,z_4)&\equiv& B^{\nu\bar\nu}(x_i,0)+B^{\nu\bar\nu}(0,z_4)-B^{\nu\bar\nu}(0,0)-B^{\nu\bar\nu}(x_i,z_4)\,,
\end{eqnarray}
with
\begin{eqnarray}
B^{\mu\bar\mu}(x_i,y_j) &=& \frac{1}{4}\left[U(x_i,y_j)+\frac{x_i y_j}{4}U(x_i,y_j) - 2 x_i y_j \tilde U(x_i,y_j) \right]\,, \label{Bmumu}\\
B^{\nu\bar\nu}(x_i,y_j) &=& \frac{1}{4}\left[U(x_i,y_j)+\frac{x_i y_j}{16}U(x_i,y_j) +\frac{x_i y_j}{2} \tilde U(x_i,y_j) \right]\,, \label{Bnunu}
\end{eqnarray}
and
\begin{eqnarray}
U(x_1,x_2)     &=& \frac{x_1^2 \log x_1}{(x_1-x_2) (1-x_1)^2} + \frac{x_2^2 \log x_2}{(x_2-x_1) (1-x_2)^2} + \frac{1}{(1-x_1)(1-x_2)}\,,\\
\tilde U(x_1,x_2) &=& \frac{x_1 \log x_1}{(x_1-x_2) (1-x_1)^2} + \frac{x_2 \log x_2}{(x_2-x_1) (1-x_2)^2} + \frac{1}{(1-x_1)(1-x_2)}\,.
\end{eqnarray}
We also encounter the function
\begin{eqnarray}\label{eq:P}
P(x_i,y_j) &\equiv& \frac{1}{(1-x_i)(1-y_j)} \left(1-\frac{7}{4} x_i
y_j\right) +\frac{x_i^2 \log x_i}{(x_i - y_j) (1-x_i)^2} \left( 1- 2
y_j + \frac{x_i y_j}{4} \right) \nonumber\\
&&  -\frac{y_j^2 \log y_j}{(x_i - y_j) (1-y_j)^2} \left( 1- 2
x_i + \frac{x_i y_j}{4} \right)\,.
\end{eqnarray}
Finally, the result for the anomalous magnetic moment of the muon is expressed in terms of
\begin{equation}
L_2(x)=\frac{1}{6(1-x)^4}\left(-10+43x-78x^2+49x^3-4x^4-18x^3\log x\right)\,.
\end{equation}

\end{appendix}

\providecommand{\href}[2]{#2}\begingroup\raggedright\endgroup
\end{document}